\documentclass[11pt]{article}
\usepackage{graphicx}
\usepackage{amsmath}
\usepackage{amssymb}
\usepackage{amsxtra}
\usepackage{changebar}
\usepackage{placeins}
\newcommand{\bc}{\begin{center}}
\newcommand{\ec}{\end{center}}
\newcommand{\bd}{\begin{displaymath}}
\newcommand{\ed}{\end{displaymath}}
\newcommand{\be}{\begin{equation}}
\newcommand{\ee}{\end{equation}}
\newcommand{\ba}{\begin{array}}
\newcommand{\ea}{\end{array}}
\newcommand{\bea}{\begin{eqnarray}}
\newcommand{\eea}{\end{eqnarray}}
\newcommand{\bt}{\begin{tabular}}
\newcommand{\et}{\end{tabular}}

\newcommand{\bp}{\begin{picture}}
\newcommand{\ep}{\end{picture}}
\newcommand{\bfi}{\begin{figure}}
\newcommand{\efi}{\end{figure}}
\begin{document}

\title{{\huge \bf F(750), We Miss You\\
as Bound State of 6 Top and 6 Anti  top}}

\author{
H.B.~Nielsen
\footnote{\large\, hbech@nbi.dk, 
hbechnbi@gmail.com}
\\[5mm]
\itshape{
The Niels Bohr Institute, Copenhagen,
Denmark}\\
}

\date{}

\maketitle

\begin{abstract}
We collect and estimate support 
for our long speculated 
``multiple point principle''
\cite{5mp,6mp,7mp,8mp,9mp,10mp,
11mp,12mp,13mp,14mp,15mp,16mp,
17mp, deriving, Kawana1, Kawana2, Kawana3, Kawana4, Kawana5, Kawana6} 
saying that there should be 
several vacua all having (compared
to the scales of high energy 
physics) very low energy densities.
In pure Standard Model we suggest 
there being three by ``multiple
point principle'' low energy 
density vacua, ``present'', 
``condensate'' and ``high field''
vacuum. We fit the mass of the in 
our picture since long speculated
bound state \cite{1nbs,2nbs,3nbs,4nbs,5nbs,
6nbs,7nbs,8nbs,9nbs,10nbs,11nbs,12nbs,
13nbs,14nbs,LNvacuumstability} of six top 
and six 
anti top quarks in three quite 
{\em independent ways} and get 
remarkably within our crude 
accuracy the {\em same} mass in 
all three fits! The new point 
of the present article is to 
estimate the bound state mass in 
what we could call a bag model 
estimation. The two other fits,
which we review, obtain the 
mass of the bound state by 
fitting to the multiple point principle  
prediction of degenerate vacua.

Our remarkable agreement of our
three mass-fits can be 
interpreted to mean, that we have 
calculated at the end the energy 
densities of the two extra 
speculated vacua and found that 
they are indeed very small!.
Unfortunately the recently much
discussed statistical fluctuation 
peak $F(750)$ \cite{1,2,3,4} has now been 
revealed to be just a fluctuation,
very accidentally matches our fitted mass 
of the bound state 
remarkably well with the mass 
of this fluctuation 750 GeV. 
\end{abstract}

\section{Introduction}
We have long worked on the 
speculation, that six top and six 
anti top quarks due to mainly the 
rather large value of the 
top-yukawa coupling $g_t$ and thus 
to Higgs boson exchange gets 
bound so strongly to each other, 
that a bound state with a mass 
appreciably lower than the sum 
of the masses of 12 top-quaks 
is formed\cite{mass,1nbs,2nbs,
3nbs,4nbs,5nbs,
6nbs,7nbs,8nbs,9nbs,10nbs,
11nbs,12nbs,
13nbs,14nbs,LNvacuumstability, 
FN750}. In the present article 
we shall put forward an attempt
to estimate the mass of this 
bound state by setting up a kind 
of bag-model ansatz for the 
bund state system. The ``bag'' 
here denotes a region in space, 
where the Higgs field is equal 
to zero, so that the mass of the 
quarks, e.g. the top quark, is
also zero there. Thus such a bag filled
with top and anti top quarks 
can make up a bound state, that
is identified as our long 
speculated one. We consider it
of great importance to estimate 
the mass of this bound state, 
not only because we hope, that 
LHC or some further accelerator 
might find it some day, but also
because, if it is as expected 
strongly bound, it is expected to 
function as approximately a new 
elementary particle giving rise 
to loop diagrams, that can give 
various corrections. Most 
important for the trustability 
of our long speculated picture 
being based on what we call the 
``multiple point principle'' 
saying, that there are several 
vacua with very small energy 
density \footnote{Strictly speaking the 
multiple point principle just tells that 
there are several 
vacua with the {\em same} energy density. 
But if you instead say that there are 
several vacua with 
{\em very small} energy density, 
you formally make the mystery of 
the smallness of the cosmological 
constant become formally a part 
of the 
assumption of our multiple point 
principle. For this point we thank 
L. Susskind.}   is, that calculation of 
the energy density of candidates 
for such alternative vacua is 
sensitive to the mass of our 
bound state. In our speculation 
we have in the pure Standard 
Model - but we have many 
applications also to extensions 
of the Standard Model with more 
speculated vacua - just 
three vacua, which we may call
``present'', ``condensate'' and 
``high field''. The ``present'' 
stands for the usual vacuum, in 
which we so to speak live, while
``condensate'' denotes a state in
which space is filled up with a 
smooth density of the bound 
states; they may possibly form 
a bose-condensate, but at least
they  
should be present with some 
density and interact with each 
other. In a work with C. D. 
Froggatt we approximated the 
configuration of the bound state 
distribution in the state making 
up the ``condensate'' vacuum     
by the configuration of carbon 
atoms in a diamond chrystal.
Whether the interaction between 
the bound states present in such 
a ``condensate''-vacuum can 
just compensate the 
Eintein-energy(i.e. the mass) of
the bound states, is presumably 
rather much dependent on this 
mass. Thus by a rather round 
about way Froggatt and I 
obtained a mass prediction 
285 GeV for the bound state 
\cite{Tunguska}. More directly
but basically by the same method 
I estimate\cite{mass} a mass 
of $4 m_t= 690 GeV$ from this 
requirement that the ``condensate''
vacuum be degenerate with the 
`present'' one. Also with 
Das and 
Laperashvili
\cite{LNvacuumstability} we 
estimated a rather small 
correction to the Higgs mass, that 
should correspond to the second 
minimum in the Higgs field 
effective potential - 
 what we called 
``high field''vacuum - should just 
touch zero, which is equivalent 
to the ``high field'' vacuum 
having zero(small) energy density.
Without any correction from the
bound state the Higgs mass, that 
just makes this boarderline 
stability, is\cite{Deg} 129.4
GeV, but we can assuming a 
bound state mass in the 800 GeV 
range\cite{LNvacuumstability}
obtain a correction making the 
experimental Higgs mass 125 GeV
compatible with the degeneracy 
of the ``high field'' and the 
``present'' vacua.

The result of the present article 
from calculating the bag-model
mass estimate is, that this 
estimate comes after all the 
corrections to be almost 
unexpectedly close to just the 
value that is needed for 
arranging the two alternaitve 
vacua, the ``condensate'' and the 
``high field'' one to be 
degenerate with the ``present'' 
vacuum. This would mean a 
calculational confirmation of 
our long speculated ``Multiple
Point Principle''. So we could 
claim, that if the calculations 
including the present mass 
estimate and the previous works 
concerning the degeneracies of 
the 
two alternative vacua got 
confirmed, 
to have calculated, that this 
``Multiple Point Principle'' were
simply 
true (by calculation) for 
the three vacua proposed to 
be possibilities in a picture 
of pure Standard Model. 
If as we claim crudely below 
the three vacua are indeed 
degenerate, it would not only 
imply that we established our 
- one could say new law of 
nature - the ``Multiple Point 
Principle'', but also that we 
would need the Standard Model 
to work with sufficient accuracy 
almost all the way to the Planck 
scale. At least possible 
disturbances of the pure Standard 
Model by for instance see-saw 
neutrinoes or super symmetry 
or whatever should be so small,
when counted in the Higgs 
effective potential, that it 
would only change our 
calculations negligebly.

We must admit that it must be just
a mysterious accident, that the 
average mass of our three 
remarkably coinciding masses 
happen to be very close to that mass, 
which were seen as a digamma 
$F(750)$ fluctuation\cite{1,2,3,4,Dias} 
- only 
revealed to be a fluctuation 
quite recently\cite{killingICHEP} -.

It has all the time been so, that
we have presented our bound state
story together with our 
``Multiple Point Principle''
\cite{5mp,6mp,7mp,8mp,9mp,10mp,
11mp,12mp,13mp,14mp,15mp,16mp,
17mp,
deriving, Kawana1, Kawana2, Kawana3, Kawana4, Kawana5, Kawana6}.

Historically we - Don Bennett and I at 
first - invented this 
priciple\cite{5mp,DonThesis,6mp,7mp} in 
order to justify a model, in which
we fitted fine structure 
constants by means multiple 
points in phase diagrams for 
lattice gauge theories. This were
in a model with each family of fermions having its own family of 
gauge particles (Anti GUT).       
\cite{AntiGUT, AntiGUTVolovik, AGUTold1, 
AGUTold2, AGUTold3}

We shall understand the Multiple 
Point Principle as a principle, 
that delivers restrictions 
between the coupling constants 
and mass parameters (the bare 
couplings or renormalized ones does 
not matter so much, the restrictions are 
just slightly different),
namely so as to make the zero 
energy densities for the vacua.

By providing such 
restrictions among coupling constants
it has 
the chanse to serve   
as a candidate for 
solving some fine tuning 
problems\cite{6nbs,
  7nbs}. 
We could also say that
this ``Multiple Point Principle''
 means, that 
the universe-vacuum is just at 
some 
multiple point, where several 
phases 
can coexist, much like  one at 
the triple 
point for water has coexistence 
of ice, fluid water, and vapor 
for a common 
set of intensive variables, 
pressure and 
temperature. There may be 
{\em no real 
good derivation} or argument for 
our 
``multiple point principle'' in 
spite of 
the fact, that we have published 
some attempts to derive this principle
\cite{5mp,deriving, Stillits, Kawana1, Kawana2, Kawana3, Kawana4, Kawana5, 
Kawana6},
but all such arguments would
have to involve some influence 
of the 
future on the past, at least 
on the 
coupling constants, and that 
would make 
all
such derivations of MPP 
suspicious.
The reader should rather take 
some 
previous works - even 
{\em pre}diction(s) - 
as well as the results of the 
present 
work as {\em empirical} 
evindence for 
this {\em new law of nature}, the 
``multiple 
point principle''(see also 
\cite{fraradio}). 

Our main picture, the consistency 
of which we suggest should lead
to the belief in it, consists of 
the following three ingredients
or assumptions:

\begin{itemize}
\item{1.} There is very strongly bound 
bound state of 6 top + 6 anti top quarks
(very strongly here means binding energy 
not small compared to mass-energy of the
constituents.), 
\item{2.} our ``Multiple Point Principle''
saying that there are 
some {\em different} 
ground states (=vacua) of the quantum 
field theory, all with almost zero 
energy density(=cosmological constant), 
say in the Standard Model 3.
Here the ``almost zero'' means that the 
energy density of these - actually 3 
relevant  -
different vacua are of the order of 
the energy density as determined 
astronomically to be of the order of 
3/4 of the total energy density in the 
universe.  
(Really we formulate and use slightly 
different versions - and especially 
different degrees of accuracy - of 
``Multiple Point Principle'' and for 
instance take it to mean that there are 
several vacua with the {\em same} energy 
density - and not as suggested for the 
present article that the energy densities 
are all small just, see the footnote 
above - and then we even had 
an argument for how big the energy 
density of these vacua should be by 
using an almost supersymmetric vacua 
as one of the vacua in the flock of 
degenerate vacua.\cite{15mp,16mp,17mp}.)   
\item{3.} Pure Standard Model is 
for our purpose all there is, i.e.
no new physics should be strong 
enough to disturb severely our 
calculations. So if our paper 
is sufficiently convincing in its 
consistency of the mass of the 
bound state, it should put some 
limit on how much new physics 
could be allowed. Only above 
$10^{18} GeV$ strong new physics 
can be allowed in our picture. 
Below it should at least not 
severly influence the running 
by renorm group of the Higgs self-coupling $\lambda$

  \end{itemize}  
  
The development of the present 
article is a calculate/estimate of the
mass of the bound state from the 
binding of the constituents 
mainly by Higgs exchange, but 
correction from exchange of 
gluons and of W's and Z's must 
be included.  
 
Our estimate or calculation 
is based on an approximate
picture of the bound state of 
the 6 top 
+ the 6 anti top quarks as 
consisting of 
a core or bag, in which the 
normal vacuum 
Higgs field is suspended, 
surrounded by a 
region around, in which the top 
and anti 
top, which are effectively 
massless inside 
this core or bag, tunnel out 
some disitance 
into the surrounding ``normal vacuum'', 
with 
the usual Higgs expectation value 
$<\phi_H > = 246 \; GeV$. 

However, such a picture is far 
too 
crude and needs a series of 
improvements
to become more accurate:

\begin{itemize}
\item{{\bf Tunnelling}} We must 
take into 
account, that the top and anti 
top quarks 
in the ``bag'',the region, where 
the 
Higgs field is zero, virtually 
tunnel 
out of this bag at the surface, 
and thus 
in reality are spread over a 
region, which 
includes a {\em rim} around the bag. 
The 
extend of this rim is of the 
order 
of magnitude of the inverse top 
quark mass
$1/m_t$. In a somewhat ad hoc 
way we 
tune in the precise width of 
this rim. As a first pedagogical 
excercise we tune in  
so as to ensure, that in the 
limit of zero radius of the bag 
the mass of the bound 
state 
resulting goes to the collected 
mass of 
the constituents.
 In the section \ref{changed} we improve 
on
this proceedure.
\item{{\bf Gluons}} Although one 
may expect 
that somehow this bag-model 
calculation
may take into account the binding due to
{\em Higgs exchange}, of course 
the effect
of gluon exchange cannot 
possibly have 
been included into that 
calculation. So 
a correction to include the 
effect 
of the exchange of gluons 
between the 
constituent quark and anti quarks has 
to be performed extra.

\item{{\bf Eaten}} Similarly the 
exchange 
of other components of the 
Higgs field 
than the ``radial'' components 
parallelel
to the vacuum expectation value 
must also 
be taken into account as a 
correction. 
Indeed these other components 
appear 
in the Standard Model 
essentially as the 
longitudinal components of the 
weak gauge 
bosons W's and Z. They have so to 
speak been ``eaten'' by these 
gauge 
particles. So in reality it is 
the 
exchange of the weak gauge 
particles, 
for which it is needed to correct.     
\end{itemize}  

The main point of the present 
article 
is, that this estimate of the 
mass of the 
bound state of the the 6 top + 
6 anti 
tops from bag-model 
using the parameters of Higgs 
interactions
as known phenomenologically is in a 
very similar range as the earlier 
estimates of this bound state 
mass based 
very strongly on the assumption 
of 
``Multiple Point Principle'' that there 
shall be several with the present one 
degenerate vacua.  This agreement 
namely means a test of the 
``Multiple 
Point Principle''. When the 
mass, we shall 
obtain in the present article, 
namely 
fit the needs for the validity of the 
Multiple Point Principle, then we 
are in reality confirming this 
principle.

Although thus it is the main 
point to 
suggest, that there is some 
numerical 
evidence for the Multiple Point 
Principle,
our result of course has the 
consequence 
that we predict there to be a 
bound state 
with the mass range resulting. 
Within the
uncertainties in our estimates 
this mass 
range would have fitted well 
with the 
by now  essentially dead 
F(750). 
So we predict that one shall
find a new particle in LHC with the decay 
branching roughly as described 
in our 
article \cite{FN750}. Our 
expectations
for the production rate are 
uncertain,
but would have fitted well, if 
one could already have seen the 
particle.
Although it is also well 
possible that 
one should not yet have seen 
this bound 
state, it is so that at much 
improved
LHC data it should show up in the 
future.   

In the following section 
\ref{boundMPP}
we shall review our model of 
there 
existing an exceedingly 
strongly bound 
system of 6 top + 6 anti top 
quarks, 
and of our 
``multiple point principle'' 
fine tuning the coupling 
constants, so 
that for instance a condensate 
of the 
bound state can fill the vacuum 
and 
cause a ``new vacuum''
(``condensate'' vacuum) with the 
energy density just finetuned 
to be again 
remarkably small, of the same 
order as 
say the astronomical observation 
of the 
energy density(= cosmological 
constant)
of the vacuum, in which we live.
(This astronomically observed 
cosmological
constant is quite negligible 
compared 
to the energy densities of any 
significance for high energy 
physics 
parameters such as the bound state 
mass or
the Higgs mass). 
A subsection \ref{MPP} of this 
section 
\ref{boundMPP} is assigned to 
our ``new law of nature'',the 
``Multiple 
point
principle''.

In the following section 
\ref{basic} 
we set up the basic bag-like 
model, and 
in the subsection \ref{adjust} 
we adjust
a parameter $a$ so as to obtain 
at least 
for the case of a very small 
bag the mass
of the bound state going to the 
sum of 
the masses of the constituents, 
as must be the case. (Really it should 
be even smaller to be discussed later
in \ref{changed}).
In the subsection 
\ref{bagconstant} 
we evaluate the energy density 
required 
to put the Higgs field to zero 
rather 
than to the usual 246 GeV.

In section \ref{changed} we improve on the 
adjustment which we made in subsection 
\ref{adjust} by recognizing that the bag 
first develop when the coupling has become 
quite strong.
In section \ref{corrections} we 
then seek 
to correct the first crude 
result for the 
mass of the bound state, first 
of all
for the exchange of weak gauge 
bosons 
- especially the zero helicity 
components 
called ``eaten Higgses'' - and 
the 
effect of gluon exchange.

In section \ref{worry} we seek 
to convince ourselves and the 
reader, that the way we corrected 
for the ``eaten '' Higgs exchange
is not completely crazy 
physically.    
In section \ref{conclusion} we 
review 
and comment our result. 
      
\section{Bound State 
Picture and 
``Multiple 
Point Principle''}
\label{boundMPP}
The crucial suggestion behind 
our bound 
state model of 6 tops + 6 anti 
tops 
is, that, since Higgs 
exchange 
like any other even order 
tensor particle 
exchange delivers attracktion 
between 
top and top, or top and anti top,
 or 
anti top and anti top as well, we 
get stronger and stronger 
binding between 
the 
top and anti top quarks the more 
of them 
we imagine brought
together. It is because the top 
and anti 
top are the strongest binding 
quarks, that
this type of binding becomes 
most relevant
for the top and anti top. Now, 
however,
the quarks are fermions and 
thus you 
cannot just unlimmited clump
arbitrarily many, e.g. top 
quarks, 
together.
Since the top quark has a color 
degree
of freedom taking three values,
say: red, 
blue,
and yellow, and a spin degree 
of freedom,
that can be up and down, one can 
bring up to
{\em 3*2=6 top quarks into the  
same 
orbital
state}, but because of 
fermi-statistics 
no more. So there can in a 
single orbital (meaning here a 
basis {\em wave function} for the 
{\em positional}($\sim$ momentum) 
degrees of freedom)  
state be just up to 6 top + 
6 anti top.
Thereby a {\em closed shell} is 
so to 
speak 
formed (in the nuclear physics 
sense).
In the zero Higgs mass 
approximation,
which will be effectively valid, 
when the 
size of the bound state - the 
radius - multiplied by the 
effective 
Higgs mass is small, the 
attracktion 
between the top-quarks or between 
tops and anti tops is quite 
analogous 
to that between an atomic 
nucleus and an
electron. So  we can for first 
orientation use the terminology 
from 
the quantum mechanical 
description 
of atomic physics. Approximating 
the 
bound state, that we suggest to 
be 
possible to form from  
6 top + 6 anti top 
by thinking
of each top or anti top going 
around a 
collected object formed from 
the other 
11 quarks, we can talk about 
different 
``orbits'' in the atomic 
terminology 
of a main quantum number $n$ 
taking 
positive integer values and 
further $l$
(the orbital angular momentum 
magnitude 
being $\sqrt{l(l+1)}$) 
and $m$(the angular momentum 
around the 
quantization axis). As in 
atomic physics 
the 
particles in the n=1 orbit are 
bound strongest. This is analogously to 
the 
helium atom having especially 
high 
excitation energies($\sim$ being 
especially strongly bound relatively to 
neighbors in the Mandelejev 
system); we, however,
 have 
because 
of the color factor 3 and both 
quark 
and anti quark an especially 
stable(strongly bound) 
system being a bound state of 
6 top 
and 6 anti top quarks. 

Whether the binding of such a 
system
of 6 top + 6 anti top now is 
sufficiently 
strong to even bind 
to form a resonance, let alone 
with the 
rather small mass 
compared to 
the collective mass of 
6 top + 6 anti top,
 12 $m_t$ = 12* 173 GeV = 
2076 GeV, is 
controversial\cite{Kuch1,Kuch2,
Kuch3}.
However, we think 
ourselves\cite{4nbs, remarkable},
that making use of a long 
series of 
corrections, especially also 
exchange 
of the other three components 
of the 
Higgs than the as simple particle 
observed component, we can 
stretch 
the uncertanties in the 
calculation 
so far as to allow  a (very) light bound 
state to be possibly 
formed\cite{4nbs}. 
These other components
of the Higgs are really present 
in the 
Standard model as W's and Z 
longitudinal 
components. We call them 
``eaten Higgses'',
but really of course it just 
means to 
include (longitudinal components of) 
weak gauge particle Z 
and W  
exchanges.(We believe these longitudinal 
componets to be actually  more 
important than the transverse components 
in the binding, but of course for high 
accuracy even the transverse components 
should be included). 
 
It is important for our hope, 
that the 
bound state can indeed bind so 
strongly, 
that it get so tightly bound, 
that the 
strong Higgs fields inside the 
bound state
even can modify the effective 
mass of the 
Higgs significantly there. 
{\em We}\cite{remarkable} 
estimated that 
a top-Yukawa coupling of $g_t 
=1.02 \pm 14 \%$ would be just 
sufficient
to bind an extremely light 
bound state 
of the 6 top and 6 anti top, and 
would 
match with the experimental 
top-Yukawa
$g_t =0.935$. But Shuryac et al. 
\cite{Kuch1, Kuch2,Kuch3}
find, that due to the high 
Higgs mass, it 
cannot bind at all for the 
experimental 
value of $g_t$.
     
\subsection{MPP}
\label{MPP}
The whole speculation about our 
bound 
state of 6 top + 6 anti top is 
a priori 
rather much taken out of the air 
by itself.
However, we have all the time 
proposed 
it only connected with another 
speculation,
the ``Multiple Point Principle''. 
This is, you
could say, a wild guess about 
simplifying or unifying
the fine tuning problems of the 
Standard Model. In order to 
formulate 
just the cosmological constant 
problem\cite{Weinberg}
about, why the cosmological 
constant(= 
the vacuum energy density) 
compared to 
say Planck scale dimensional 
expectations
is so enormously small, one 
needs an 
{\em assumption} of the form 
``The energy
density of vacuum is extremely 
small!''
Now you could look at the 
``Multiple Point 
Principle'' as an {\em extension}
- or putting into ``plural'' - of 
this anyway
needed assumption, {\em without} really 
{\em complicating} it severely: 
``Several 
vacua have extremely small energy 
densities!''. We almost just 
have 
put the anyway needed 
assumption into 
``plural'', or
changed the ``quantor'' from  
``The physical vacuum...'' to 
``Several
vacua...''.

Now the real supporting point 
for this 
principle is, that although it 
is not 
unneccesarily 
complicated, it  is the one, 
which Colin 
Froggatt and 
I managed to use to make 
historically\cite{9mp} in 1996, 
long before the Higgs 
particle were found, 
a 
{\em pre}diction of
the Higgs mass of 135 GeV(or 130 GeV) 
$\pm$ 10 GeV .
Now our prediction using the same
Multiple Point Principle would be 
129.4 GeV\cite{Deg} but with a 
much 
smaller 
uncertainty, comparable to the 
experimental 
uncertainty of a few hundred 
MeV. So 
at first it then looks, that 
while our 
original {\em pre}diction agreed 
perfectly
within errors, and 
the Multiple Point Principle were 
perfectly right, it is today 
deviating
of the order of three standard 
deviations (some uncertainty 
comes from the mass of the top 
quark, 
which goes strongly into the 
Higgs mass prediction) 
from matching experiment.
This formal disagreement of the 
theoretical prediction 
actually came  in spite of, 
that the 
better 
calculations and better
top mass moved our 
prediction {\em closer}
 to 
the experimental
value 125 GeV during the time we 
had 
{\em pre}dited it. It is of  
course 
only 
possible, that in spite of this 
development the agreement 
relative to the uncertainty 
could become worse, because the 
uncertainties in 
calculation and top and Higgs 
masses 
went down even faster. However, 
L.V. 
Laperashvili, 
C. Das and myself
\cite{LNvacuumstability}
 found, that 
the existence of the bound state 
of the 6 top + 6 anti top would 
make a 
{\em little theoretical 
correction to 
the mass 
of the Higgs}
being predicted from the 
multiple point principle,
so that the agreement might 
indeed 
be improved to be perfect, if the 
mass of this bound state is 
appropriate. According to our 
estimates a mass about 850 GeV 
or 710 GeV is what is very 
crudely
called for. 
 

It should be stressed, that this 
successfull Higgs-mass 
{\em pre}diction 
as well as Colin D. Froggatts 
and mine\cite{remarkable, 4nbs} 
controversial argument, that the 
top-Yukawa-coupling $g_t$ in 
order to 
allow for a condensate of bound 
states of 
6 top and 6  antitop with 
energy density 
close to zero, must be close to 
the 
value $1.02 \pm 14 \%$ 
supports the ``Multiple Point 
Principle'' as being a principle 
uphold by nature. The value 
$g_t = 1.02 \pm 14 \% $ namely 
matches
with the experimentally 
determined 
Higgs Yukawa coupling 
$g_t = 0.935$. 
Really we just estimated, what 
the top-Yukawa coupling should 
be in order,
that the bound state assumed to 
exist 
of 6 top + 6 anti top should have 
exceptionally low mass. But this 
should be 
approximately 
needed to have the 
condensating particle 
have mass close to zero in 
order for there
being two degenerate vacua as 
required 
by MPP.

This means that even, if the 
theoretical 
arguments for the MPP are not 
totally 
convincing,
there is nevertheless some {\em 
empirical 
evidence} pointing in favor of 
this MPP.    And the present 
article is meant to 
provide one more such indirect  
phenomenological support for MPP.

\section{Bag-like Model}
\label{basic}
\subsection{First Charicature}
The basic idea of our 
caclculation/estimation of the 
mass of 
the hoped for bound state of 6 
top and 
6 anti top is, that this bound 
state is 
in the very crudest 
approximation a 
sphere of radius $R$, inside 
which 
the Higgs field $\phi_H$ is 
arranged to 
be fluctuating quantum 
mechanically 
around zero, rather than around 
the usual 
$246$ GeV. Inside this ball or 
bag 
of suspended Higgs field the 
quarks are 
in principle massless and our 
picture 
of the bound state then consists 
in 
this inside ball region being 
filled 
by the 6 top + 6 anti top quarks, running 
of course with their by 
Heisenberg 
uncertainty principle required 
momentum in average. Then the proceedure 
is to 
consider the energy of an ansatz with 
a given radius $R$ of the bag a function 
of this radius, and then minimize  
this energy w.r.t. to the radius. The 
minimal energy is then the first 
approximation to the mass of the bound 
state. 

For pedagogical reasons let us first 
put forward this far too crude model:

The energy for the bound state ansatz 
is in this first approximation given as 
the sum of two terms:
\begin{itemize}
\item{{\bf Bag-constant term}}
In the next to following section
\ref{bagconstant} we estimate the energy
density of a piece of space in which the 
Higgs field has been imposed to fluctuate 
around $\phi_H =0$, i.e. the effective potential
for the Higgs field at $\phi_H =0$,
\begin{equation}
V_{eff}(\phi_H = 0) =\frac{ v^2m_H^2}{8} = 
0.132 m_t^4 = 1.18 *10^8 GeV^4 
\end{equation}  
where we have used 
\begin{eqnarray}
\hbox{Normal Higgs vacuum expectation 
value  } v
= <\phi_H>&=& 
246 GeV\\
\hbox{Higgs mass  } m_H &=& 125 GeV\\
\hbox{Top quark mass  } m_t &=& 173 GeV
\end{eqnarray}
and have normalized the effective 
potential to be zero for the value taken 
on in the usual vacuum. I.e. we normalized
to $V_{eff}(v) =0$. 

Then the energy of the whole bag is 
\begin{eqnarray}
E_{bag}&=& \frac{4\pi}{3}R^3V_{eff}(0)\\
&=& \frac{4\pi*0.132m_t^4}{3}R^3 
= 0.553 m_t^4 *R^3\\
&=& 4.95 *10^8 GeV^4 *R^3 
\end{eqnarray}
\item{{\bf Fluctuation Kinetic Energy}}
If a particle is distributed evenly over 
the volume of the bag having radius $R$
the average of the square of its distance 
form the center of the bag will be
\begin{equation}
<\vec{r}^2> = \frac{\int_0^Rr^2 *r^2 dr}
{\int_0^Rr^2dr} = \frac{3}{5}R^2 
\end{equation}
and thus using Heisenberg uncertainty 
principle in the three dimensional form
\begin{equation}
<\vec{r}^2><\vec{p}^2> \ge \frac{9}{4} 
\bar{h}^2
\end{equation}
with the inequality taken as an 
{\em equality} we derive, that 
the average of the momentum 
squared of the 
particle/consituent confined in 
the bag 
must be at least 
 \begin{equation}
<\vec{p}^2> \ge \frac{9}{4}*\frac{5}{3R^2}
= \frac{15}{4}* R^{-2}.
\label{vecfluc}
\end{equation}
For an in the interior of the 
bag massless
particle one has of course 
$E_{\hbox{one particle}}= |\vec{p}|$ 
and thus a crude 
estimate of the energy on the average 
is $E_{\hbox{one particle}}\approx 
\sqrt{<\vec{p}^2>}$. Thus we get as an 
estimate for the (kinetic) energy for 12 
constituents, which are 
inside the bag a lower limit, which is 
actually crudely an estimate in very 
first approximation:
\begin{eqnarray}
E_{kin}&\approx& 12 * \sqrt{<\vec{p}^2>}\\
& \ge&  12 * \sqrt{\frac{15}{4}* R^{-2}}\\
&=&\frac{ 12 * \sqrt{\frac{15}{4}}}{R}\\
&=& \frac{\sqrt{15 *9 *4}}{R}\\
&=& \frac{23.24}{R}.\label{Ekin}
\end{eqnarray}   
\end{itemize} 

The total energy of the bound 
state model
ansatz is then the sum 
$E_{bag}+E_{kin}$ 
given by
\begin{eqnarray}
M_{ansatz} &=& E_{bag}+E_{kin}\\
&\approx&  0.553 m_t^4 *R^3 +
\frac{ 12 * \sqrt{\frac{15}{4}}}{R}.
\label{Man}
\end{eqnarray} 

Now we have to find that value of $R$ 
which gives the lowest $M_{ansatz}$ and 
that minimum value of $M_{ansatz}$ should
then approximately be the bound state mass.
In order to find this minimum we therefore
differentiate the expression (\ref{Man})
w.r.t. $R$:
\begin{eqnarray}
\frac{dM_{ansatz}(R)}{dR}&\approx&
3* 0.553 m_t^4 *R^2 -
\frac{ 12 * \sqrt{\frac{15}{4}}}{R^2}.
\end{eqnarray}
Putting this derivative to zero 
leads to 
\begin{eqnarray}
R^4 &=&
\frac{ 12 * \sqrt{\frac{15}{4}}}
{3* 0.553 m_t^4 }\\
&=&  \frac{12 *1.673}{m_t^4}\\
&=&12 *1.303*10^{-9} GeV^{-4}\\
&=&   1.564 *10^{-8} GeV^{-4}.
\end{eqnarray}
This leads to the radius giving 
the lowest mass ansatz being
\begin{eqnarray}
R &=& \left ( \frac{12 *1.673}
{m_t^4} \right )^{1/4}\\
&=& 12^{1/4} * 
1.137/m_t\\
&=& 12^{1/4} 
6.574 *10^{-3} GeV^{-1}
\end{eqnarray}

The proceedure to find the mass 
is to 
insert this value of $R$ into the 
expression (\ref{Man}), and that 
leads to 
\begin{eqnarray}
M_{ansatz}|_{min}&=&  0.553 m_t^4 *(12^{1/4} *
1.137/m_t  )^3 +
\frac{ 12 * \sqrt{\frac{15}{4}}}{12^{1/4} *
1.137/m_t }\\
&=&m_t* 12^{3/4}*(0.553*
1.137^3+\sqrt{\frac{15}{4}}/
1.137)\\
&=& m_t*12^{3/4}( 
0.813+ 
1.703
 )\\
&=& m_t*12^{3/4}*
2.516\\
&=& 12^{3/4}*
435 GeV\\
&=& 
2806 GeV
\end{eqnarray}
So the mass of the bound state is 
estimated in this first ``calculation''
as 
2806 GeV. But that is 
{\em crazy}, 
because a bund state should have at most 
the mass of the collection os constituents
\begin{equation}
\hbox{Collection of constituent 
mass}
= 12 m_t = 2076 GeV.
\label{colectedmass}
\end{equation}  
One mistake we have made, is to 
assume that 
the quarks cannot come out of 
the region(=the bag) 
in which they are massless, even 
by 
tunnelling. But that is of 
course not true;
rather one would expect the 
quarks to 
tunnel outside the bag over  a 
length
of the order of a top-quark 
compton wave 
length $1/m_t$. Thay would 
effectively 
increase the radius to be used 
for 
estimating the kinetic energy 
term 
$E_{kin}$ by an extra amount of 
this order of magnitude. Really 
it means that
we must imagine a {\em rim} 
around the bag 
of thickness of this order $\sim 1/m_t$.
 
But of course the also the Higgs field does {\em not} jump at the bag-surface
as we used at first.
\subsection{Adjusting the Tunnelling 
Around the Bag}
\label{adjust}
The proceedure proposed in the 
present 
article is to seek to make, what 
we could call a semi-empirical 
formula
for the mass of the bound state, 
by 
adjusting an ad hoc coefficient 
$a$ 
to make the result at least so 
sensible
as to give a sensible result in 
the limit of a small bag 
- i.e. small $R$ - which 
corresponds to 
no effect of the Higgs field and 
thus 
the Higgs exchange except the 
usual 
background giving the masses to 
the quarks. 
We shall below in section 
\ref{changed} modify this point of view,
and indeed rather seek to 
estimate, that when the bag at all
begins to show up in the 
structure of the bound state, 
there is already an appreciable 
binding, and the mass should 
 already be rather 3/4 *
``the mass of the constituents''.
However, to obtain first a 
relativly simple orientation of 
the order of magnitudes for the 
mass obtinable, we shall now 
calculate with the simple 
assumption, that the binding just 
begins at zero radius $R$, 
although this is not true. It 
is at least better than the 
crazy result of getting a bigger 
mass than the collected 
constituent mass. 

The more sensible result 
in this 
small $R$ limit, which we expect
- if it were not for that there 
could for low binding be no bag
at all, but
rather only the rim -  
is 
that the ground state mass must 
equal 
the sum of the masses of the 
constituents
(\ref{colectedmass}). We choose 
to use as the adjustable 
parameter the 
thickness of the rim by putting 
the 
correction of the $R$ to be 
used for the 
$E_{kin}$ via the Heisenberg uncertainty
to be 
\begin{equation}
R \rightarrow R + \frac{a}{m_t}.
\end{equation}    
That is to say we shall take 
instead of 
(\ref{Ekin}) the kietic energy to be 
\begin{eqnarray}
E_{kin}&\approx& 12 * \sqrt{<\vec{p}^2>_{mod}}\\
& \ge&  12 * \sqrt{\frac{15}{4}* 
(R+\frac{a}{m_t})^{-2}}\\
&=&\frac{ 12 * \sqrt{\frac{15}{4}}}{R+\frac{a}{m_t}}\\
&=& \frac{\sqrt{15 *9 *4}}{R+\frac{a}{m_t}}\\
&=& \frac{23.24}
{R+ \frac{a}{m_t}},
\label{Ekinmod}
\end{eqnarray}
and then we must adjust $a$ so as to 
make the whole mass in the limit 
$R\rightarrow 0$ - but that means 
effectively in this limit the 
$E_{kin}$
term alone (because then $E_{bag} 
\rightarrow 0$) - to be the 
collected consistituent mass
$12 m_t=2076 GeV$. Thus we see,
 that we must choose 
$a =\sqrt{\frac{15}{4}}$.
   
With this adjustment we obtain 
for the 
ansatz energy of the bag energy 
 plus the kinetic energy
\begin{eqnarray}
M_{\hbox{ansatz $a=\sqrt{15/4}$ }}&=&
\frac{4\pi}{3}R^3V_{eff}(0)+
 12 * \sqrt{<\vec{p}^2>_{mod}}\\
& =& 0.553 m_t^4 *R^3+
 12 * \sqrt{\frac{15}{4}* 
(R+\frac{\sqrt{15/4}}{m_t})^{-2}}\\
&=& 0.553 m_t^4 *R^3+
\frac{ 12 * \sqrt{\frac{15}{4}}}{R+\frac{\sqrt{15/4}}{m_t}}\\
&=& 0.553 m_t^4 *R^3+ \frac{23.24}{R+ \frac{\sqrt{15/4}}{m_t}}.
\label{mod}
\end{eqnarray}

Now we must of course as before 
differentiate this expression 
(\ref{mod})
w.r.t. $R$:
\begin{eqnarray}
\frac{dM_{\hbox{ansatz $a=\sqrt{15/4}$}}(R)}{dR}
&=& 3* 0.553 m_t^4 *R^2- \frac{23.24}
{(R+ \frac{\sqrt{15/4}}{m_t})^2}.
\end{eqnarray}
Putting the derivative to zero 
leads to the equation for the 
$R$ giving the minimal value 
for the ansatz energy:
\begin{eqnarray}
R^2 \left (R+\frac{\sqrt{{15}/{4}}}{m_t}
\right )^2&=&\frac{23.24}{3 *0.553m_t^4}
=14.00/m_t^4
\end{eqnarray}
 Writting explicitely the 12 
used as the 
number of constituents this 
relation takes the form
\begin{eqnarray}
R^2 \left (R+\frac{\sqrt{{15}/{4}}}{m_t}
\right )^2&=&\frac{12*\sqrt{15/4}}{3 *0.553m_t^4}
=12*1.167/m_t^4.\label{relation}
\end{eqnarray}
Taking the square root of both 
sides 
of this equation 
(\ref{relation}) leads to
\begin{eqnarray}
R \left (R+\frac{\sqrt{{15}/{4}}}
{m_t}
\right )&=&
\sqrt{\frac{12*\sqrt{15/4}}
{3 *0.553m_t^4}}\\
&=&\sqrt{12}*1.080/m_t^2.
\label{sqrt}
\end{eqnarray}
This equation is a second order 
equation 
in $R$ 
\begin{eqnarray}
R^2+\frac{\sqrt{{15}/{4}}}
{m_t}*R -
 \sqrt{12}*1.080/m_t^2& =&0
\label{sqrtos}
\end{eqnarray}
and is solved by
\begin{eqnarray}
R &=&- \frac{\sqrt{{15}/{4}}}
{2m_t}
\pm \sqrt{\frac{15}{16m_t^2}+
\sqrt{12}*1.080/m_t^2 }   
\end{eqnarray}
The radius including the 
tunnelling 
rim around the genuine bag $R+
\frac{\sqrt{15/4}}{m_t}$ thus is
\begin{eqnarray}
R+\frac{\sqrt{15/4}}{m_t}  
&=& \frac{\sqrt{{15}/{4}}}{2m_t}
\pm \sqrt{\frac{15}{16m_t^2}+
\sqrt{12}*1.080/m_t^2 },   
\end{eqnarray}
and numerically we get in units 
of top-quark masses $m_t$ for 
the physical solution having 
the plus sign in front of the 
square root
\begin{eqnarray}
R+\frac{\sqrt{15/4}}{m_t}  &=& 
0.968/m_t + \sqrt{0.9375+ \sqrt{12}*1.080}
/m_t\\
&=&0.968/m_t + \sqrt{4.678}/m_t\\
&=&0.968 m_t^{-1}+ 2.163m_t^{-1}\\
&=& 3.131 m_t^{-1}\\
\hbox{while $R$ itself is:} &&\\
R &=&-0.968m_t^{-1}+2.163m_t^{-1}\\
&=& 1.195 m_t^{-1}.     
\end{eqnarray}
Using these results for the 
radius 
with and without the rim 
inclusion we
obtain the mass of the ansatz 
bound state
\begin{eqnarray}
M_{\hbox{ansatz $a=\sqrt{15/4}$ }}|_{min}&=&
 0.553 m_t^4 *(1.195m_t^{-1})^3+
\frac{ 12 * \sqrt{\frac{15}{4}}} 
{3.131 m_t^{-1}}\\
&=& 0.553*1.195^3 m_t+
\frac{ 12 * \sqrt{\frac{15}{4}}}{3.131}
m_t\\
&=& 0.944m_t +12*0.618m_t\\
&=& 0.944m_t+7.422 m_t\\
  &=& 8.366 m_t\\
&=& 1447 GeV
\label{modmin}
\end{eqnarray}

This means that our ad hoc 
modification 
of the bag with rim of 
tunnelling precense
of the quarks leads to the 
mass 1447 GeV 
for the bound state having 
used the 
assumption also of effective 
equality in 
the Heisenberg uncertainty 
relation.
This is already an impressive 
binding, but
we have got this value only 
including 
the effect of binding from the 
exchange 
of what we call the ``radial'' 
Higgs, 
namely the experimentally 
observed 
Higgs component. But we have got this 
mass 
1447 GeV not using the gluon 
exchange force
nor what we call 
``eaten Higgses'' 
meaning essentially, that we 
did not 
include the effect of weak 
gauge boson 
exchanges.

\subsection{Bag Energy Density}
\label{bagconstant}
For the very low field region, 
which 
is relevant for estimating 
the ``bag 
constant'' $V_{eff}(0)$, i.e. the 
energy density price for 
removing the 
Higgs field $\phi_H$, the 
renormalization
group running is only of minor 
significance, and calling the 
expectation 
value of the Higgs field in the 
present
vacuum for $v$ we may write the 
effective 
potential as 
\begin{equation}
V_{eff}(\phi_H) = K(\phi_H^2 - v^2 )^2,
\label{potential} 
\end{equation} 
where then mass square of the 
Higgs 
particle equals the second 
derivative 
of the potential at the vacuum 
point 
$\phi_H=v$:
\begin{equation}
m_H^2 = \frac{d^2}{d\phi_H^2}
V_{eff}(\phi_H)|_{\phi_H=v}
= K *2 (\phi_H+v)^2|_{\phi_H = v}
= 8Kv^2,
\end{equation}   
and thus $K=m_H^2/(8v^2)$, so 
that the 
energy density for zero field is
\begin{equation}
V_{eff}(\phi_H =0) 
= Kv^4 =m_H^2v^2/8= 
(125GeV *246 GeV)^2/8 \\
=(1.014 m_t)^4/8 
=0.132 m_t^4= 1.1820*10^8 GeV^4 
\end{equation}

\section{Changed Point 
of View}
\label{changed}

In the above calculation  we effectively 
assumed, that the weak 
coupling limit were equal to 
the limit of the bag radius $R$ 
going to zero
But this is not true
because our bag has zero Higgs 
field in it, while there will 
in a 
realistic situation with a weak 
coupling 
be no appreciable or dominant 
region, in 
which the Higgs field is zero. 
It will in fact only be  close to zero or 
formally even negative in an 
extremely 
small region about each 
constituent due 
to the formal divergence of 
the 1/r like 
Yukawa potential. If one 
considers 
that there are several 
constituents 
fluctuating around the potential 
comming 
from the sum of these different 
constituents in a slightly 
smeared 
approximation will have no zeros 
in the 
weak coupling case at all. So 
taking a bag
with zero field as the major 
approximation
is definitely rather bad for weak 
coupling. We can thus only hope 
that this 
bag-model ansatz can be good for 
sufficiently strong coupling. But if so, 
then the adjustment of the 
``ad hoc'' 
parameter $a$ used above from the 
requirement, that in the weak 
coupling case
the energy of the bound state 
should equal
the sum of the constituent 
masses becomes 
meaningless. If our ansatz does 
not apply 
for weak coupling, then we 
should not 
use the weak coupling limit to 
adjust 
our parameters.

Let us therefore at least in 
words rather
think of doing the following:

We seek to construct an ansatz 
this way:
\begin{itemize} 
\item For weak coupling and a 
long way 
up in coupling we construct an 
``ansatz
potential'' being as well we can 
describe 
it at large $r$ (= the distance 
to the center) the 
Coulomb or bettter the Yukawa 
potential 
from all the constituents except 
for the 
one consisdered, and then at 
smaller 
distances $r$ it continuously 
gets reduced
in strength corresponding to 
that only a 
part of the constituents are at 
shorter 
distance than $r$ and thus 
contribute to 
the Yuakawa potential at that 
distance. 
Thereby we obtain for the 
averaged field
felt by a considered constituent 
from the 
other ones a potential that 
chopped down 
or flattened off as $r$ becomes 
smaller.
Because of this flattening off 
or chop 
down in magnitude the Higgs field 
describing approximate potential 
from the 
11 constituents never reaches down to zero.
\item Only as we let the 
coupling be 
stronger and stronger the in 
this way 
constructed potential and 
corresponding Higgs field will 
have this Higgs field
touch zero, first in the center 
$r=0$ 
of course. But now formally we 
would like 
for even stronger coupling to 
let the 
Higgs field  very near the 
center - i.e.
for $r$ rather small - go 
negative. But 
negative Higgs field does not 
really help
attrackting a top quark. The 
point is 
that the attracktion with a 
Higgs 
potential really comes about, 
because the 
mass of say a top quark is 
proportional
to the numerical value of the 
Higgs field.
But then because of this 
``numerical''
value making the Higgs field 
negative 
comes to function rather as a 
positive 
potential anyway. Therefore we 
are stuck by seeking to go 
further along 
by making a potential ansatz 
once we 
pass the strength at which the Higgs field
in the center $r=0$ becomes zero. 
\item Now we would get a better 
binding
by letting the Higgs field never 
go to negative values, but 
rather stay at zero even though 
the coupling goes 
further. That means that from 
some finite coupling strength 
at which the 
first ansatz Higgs field reaches 
zero 
further increase of the coupling 
suggests
to have a then larger and larger 
region 
with just zero Higgs field, the 
bag.

This means then that a bag model 
like 
ansatz becomes appropriate 
{\em only 
after we have reached a so 
strong coupling situation that 
the effective field 
describing the effect of the 11 
other 
constituents on one of them 
becomes so 
strong as to have the Higgs field reach 
zero in the midle.}   
\end{itemize} 
     
For the fitting of the parameter 
$a$ giving width of the rim in 
units of 
top-compton wave lengths the 
just given 
consideration means that we 
should not fit 
the zero-bag radius limit 
$R\rightarrow 0$
giving the bound state mass 
equal to the 
sum of the constituent masses, 
but rather the mass of the 
ansatz achieved just when the 
effective Higgs field 
describing the potential for one 
of the 
constituent just begin to touch 
zero 
at $r=0$. That is to say we must 
first 
estimate the bound state mass in 
the situation, where this 
effective Higgs field 
has not yet reached to zero, but just come 
there to touch zero.

Very crudely we can now say, 
that the 
Higgs 
field in radial direction runs 
from zero to the usual vacuum 
expectation value in 
the transition situation. Very 
crudely we might then say that 
the top mass in this 
range runs from zero to the top 
mass in the usual vacuum, which 
is what we call 
$m_t$. Then one could say that 
on the 
average the mass in this range 
is $m_t/2$.
Now though the top quark present in this 
region having such a crude half mass also
would have a kinetic energy 
larger than, if 
it were at rest. From virial 
theorem 
one usually take it that in a 
$1/r$ like 
potential has the kinetic energy 
being (minus) the half the 
potential 
one. The latter is the 
difference of 
the mass that were $m_t/2$(on average) and the usual
mass $m_t$. So we estimate in the 
transition case - wherein the 
effective 
Higgs field just reach to zero at $r=0$ -
that the bound state mass is 12 
times 
$\frac{3}{4}m_t$. This comes 
about, because
the binding is estimated to 
$\frac{1}{2}
*\frac{1}{2} m_t$ for the 
transition 
situation. 

We should then in the limit of 
$R\rightarrow 0$ rather than the 
mass 
12 $m_t$ require the mass 
$12* \frac{3}{4}
m_t$. 

This corresponds to that the rim 
is somewhat broader than we 
assumed above.

This means that now we must in 
the equation (\ref{mod}) above 
replace 
$m_t \rightarrow \frac{3}{4} m_t$ where 
it goes into giving the 
thickness of the 
rim so as to guarantee that in 
the small 
$R$ limit we have the suggested 
mass 
$\frac{3}{4}* 12 m_t$ for the 
bound state.
The value on the other hand of 
the bag 
constant is not changed of course.
\begin{eqnarray}
M_{\hbox{ansatz $a=\sqrt{15/4}$, improved }}&=&
\frac{4\pi}{3}R^3V_{eff}(0)+
 12 * \sqrt{<\vec{p}^2>_{mod}}\\
& =& 0.553 m_t^4 *R^3+
 12 * \sqrt{\frac{15}{4}* 
(R+\frac{\sqrt{15/4}}{\frac{3}{4}m_t})^{-2}}\\
&=& 0.553 m_t^4 *R^3+
\frac{ 12 * \sqrt{\frac{15}{4}}}
{R+\frac{\sqrt{15/4}}{\frac{3}{4}m_t}}\\
&=& 0.553 m_t^4 *R^3+ 
\frac{23.24}{R+ \frac{\sqrt{15/4}}
{\frac{3}{4}m_t}}.
\label{mod34}
\end{eqnarray}

Differentiating so as to seek 
the minimum
for this by the $\frac{3}{4}$ 
factor in 
the top-mass to give the rim 
thickness gives
\begin{eqnarray}
\frac{dM_{\hbox{ansatz $a=\sqrt{15/4}$ 
improved}}
(R)}{dR}
&=& 3* 0.553 m_t^4 *R^2- 
\frac{23.24}
{(R+ \frac{\sqrt{15/4}}
{\frac{3}{4}m_t})^2}.
\end{eqnarray}  
and thus the equation to 
determine this
minimum becomes 
\begin{eqnarray}
R^2 \left (R+\frac{\sqrt{{15}/{4}}}
{\frac{3}{4}m_t}
\right )^2&=&\frac{12*\sqrt{15/4}}
{3 *0.553m_t^4}
=12*1.167/m_t^4.\label{relation34}
\end{eqnarray}
The roots of the square root of 
this equations then lead to 
\begin{eqnarray}
R+\frac{\sqrt{15/4}}
{\frac{3}{4}m_t}  &=& 
0.968/(\frac{3}{4}m_t) + 
\sqrt{0.9375/(\frac{3}{4})^2+ 
\sqrt{12}*1.080}
/m_t\\ \label{solving}
&=&1.291/m_t + \sqrt{5.408}/m_t\\
&=&
1.291m_t^{-1}+ 
2.325m_t^{-1}\\
&=&
3.616m_t^{-1}\\
\hbox{while $R$ itself is:} &&\\
R &=&-1.291
m_t^{-1}+
2.325m_t^{-1}\\
&=& 
1.034m_t^{-1}.     
\end{eqnarray}
Inserting these radii into the 
mass ansatz
we obtain now with rim 
adjustment made to 
give $\frac{3}{4}$ of 12 $m_t$ 
for the 
bound state mass in the 
$R\rightarrow 0$
limit:
\begin{eqnarray}
M_{\hbox{ansatz $a=\sqrt{15/4}$ improved }}|_{min}
&=&
 0.553 m_t^4 *(
1.034m_t^{-1})^3+
\frac{ 12 * \sqrt{\frac{15}{4}}} 
{
3.616m_t^{-1}}\\
&=& 0.553*1.034^3 m_t+
\frac{ 12 * \sqrt{\frac{15}{4}}}{3.616}
m_t\\
&=& 
0.612m_t +12*0.535
m_t\\
&=& 0.612m_t+6.42 m_t\\
  &=& 
7.03m_t\\
&=& 
1216 GeV
\label{modmin34}
\end{eqnarray}

\section{Corrections}
\label{corrections}     
In addition to the simple Higgs 
exchange,
on which we have concentrated  
above, there 
is in the type of bound state, 
we consider
also gluon exchange and exchange 
of 
what we called ``eaten Higgses'',
 but 
which are really just exchange of weak 
gauge bosons with longitudinal 
polarization. In our work 
\cite{remarkable} we estimate 
the effect
of such corrections on the 
critical 
coupling value $g_{t\; crit}$ at 
which a phase transition between 
the usual 
vacuum and a vacuum with a 
condensate
or at least some filling 
with 
the bound 
states to be a factor $4^{1/4}$.indeed  we 
esitmate that the ``eaten 
Higgs'' extra Higgses function 
crudely 
in the approximate situation of 
the bound 
state being very light as if 
the $g_t$ had
been increased by the fourth 
root of 
the ratio of the new number to 
the old 
number of Higgs components, i.e. the 
fourth root of 4. Expressed for the 
square of the Yukawa coupling $g_t^2$ 
this effect of the ``eaten Higgses'' 
becomes thus a factor 2. Similarly 
one finds in our 
article\cite{remarkable}
that the inclusion of the 
gluon interaction in addition to the 
Higgs exchanges has an effect of the 
replacement:
\begin{equation}
4g_t^2 \rightarrow 4g_t^2 + 1.83.
\end{equation}      
Since $g_t^2 = 0.935^2 = 0.874$ we have 
$4 g_t^2 = 3.50$, and the addition 
of the gluon correction corresponds to
an increase of $g_t^2$ by a factor 
$ 1 + \frac{1.83}{4 * 0.874}=1+0.523$
= 1.523. 
  
Now 
we can argue, that the bag constant could be 
considered to be inversely proportional 
to the 4th power of the 
top-Yukawa-coupling $g_t$ in a notation, 
in which mass of the top $m_t$ is put to
be independent of this Yukawa coupling:  
\begin{equation}
V_{eff}(0) = 0.0323v^4 = 
\frac{0.0323*m_t^4*4}{g_t^4} 
= 0.1292*\frac
{m_t^4}{g_t^4}
\end{equation}
The collective effect crudely of both 
``eaten Higgses'' and the gluon exchanges
corresonds to a replacement of the 
square of the Yukawa coupling  by a 
factor 2 * 1.523 = 3.05.  This then 
corresponds to decreasing the bag constant 
by a factor $3.05^2=9.28$. 

This bag constant goes into the above 
as the inverse square root in the constant 
term in the quadratic equation of the 
type of the square root of equation 
(\ref{relation34}) such as say (\ref{sqrt})
or (\ref{sqrtos}).

This constant term is seen from 
(\ref{solving}) to then change as 
\begin{equation}
\sqrt{12} *
1.080m_t^{-2}\rightarrow 
\sqrt{12}*1.080m_t^{-2}*
3.05. 
\end{equation}
Thus we get for the radius including and 
not including the rim:
\begin{eqnarray}
R+\frac{\sqrt{15/4}}{\frac{3}{4}m_t}  &=& 
0.968/(\frac{3}{4}m_t) + 
\sqrt{0.9375/(\frac{3}{4})^2+ 
\sqrt{12}*1.080*
3.05}
/m_t\\ \label{solvingcor}
&=&1.291/m_t + \sqrt{13.06}
/m_t\\
&=&
1.291m_t^{-1}+ 
3.614m_t^{-1}\\
&=&
4.905m_t^{-1}\\
\hbox{while $R$ itself is:} &&\\
R &=&-1.291
m_t^{-1}+
3.614m_t^{-1}\\
&=& 
2.32m_t^{-1}.     
\end{eqnarray}

Now in calculating the mass of the bound 
state after the correction of $g_t^2$ by
the factor 
3.05 we shall 
remember that 
the bag constant going into 
the term 
$\frac{3\pi}{4}R^3 
V_{eff}(0)|_{after \; correction}$ in the bound state mass has to be 
reduced compared to the genuine 
$V_{eff}(0)$ by a factor $
3.05^2$. 

So this bag-model term becomes
\begin{equation}
\frac{4\pi}{3}V_{eff}(0)_{reduced}R^3 = 
\frac{4\pi}{3}*(1.014 m_t)^4/8/
3.05^2
(2.32m_t^{-1})^3=0.553/
3.05^2* 2.32^3m_t =
0.0594*2.32^2m_t=
0.320 m_t. 
\end{equation}
This term is rather small. 
Here we used:
 \begin{equation}
V_{eff}(\phi_H =0) = Kv^4 =m_H^2v^2/8= 
(125GeV *246 GeV)^2/8 =(1.014 m_t)^4/8 
=0.132 m_t^4= 1.1820*10^8 GeV^4 
\end{equation}
The term from the kinetic energy mainly
of the top-quarks and anti-top-quarks
is now given as
\begin{equation}
\frac{12 *\sqrt{15/4}}{4.905m_t^{-1}}=
12*0.395m_t=4.74m_t.
\end{equation} 
Together with the small bag-term this 
gives very crude corrected mass for our bound state of 6 top and 6 anti top
\begin{eqnarray}
M_{corrected \; eaten + gluons} =
0.320 m_t
+ 4.74 m_t = 
5.06 m_t= 
875 GeV.  
\end{eqnarray}
This would agree wonderfully with the mass
of the F(750) if it would 
resurect! More importantly: It also
agrees very well with estimates for the 
needed mass of the bound state to just 
make the correction to the vacuum energy 
density at the high higgs field minimum 
at $\phi_H \sim 10^{18} GeV $ so that it
corrects the present 
value\cite{LNvacuumstability} of the 
selfcoupling $\lambda(10^{18}GeV)$ from
its $-0.01\pm 0.002$ to just $0$:
\begin{eqnarray}
m_{F} &\approx& \frac{6g_tm_t}{b}
\left (
\frac{\sim 2}{\pi^2* 0.01\pm 0.002}\right)
^{1/4} \nonumber\\
\approx 2.31*173 GeV
*2.1& =&4.9 *173 GeV =850 GeV \pm 20 \%
\nonumber\\
\hbox{or without the $\sim 2$:}\;  m_{F(750)}= 2.31 *173 GeV * 1.8&=&4.1 *173 GeV =710 GeV \pm 20 \%
\nonumber  
\end{eqnarray} 

Further it fits very well with my earlier 
simple calculation 
\cite{mass}, which gave a mass of $4m_t
=692 GeV$ and which were based on assuming 
the ``condensate vacuum'' being degenerate
with the ``present'' vacuum. 
\section{Some Worries, Can we 
Trust?}
\label{worry}
Before believing the story that 
we could correct for the gluon 
exchange and eaten Higgs 
exchanges 
just by replacing the bag 
constant $V_{eff}(0)$ by a 
dramatically smaller value - 
indeed diminshed by a factor 
$3.05^2$ - 
and thus obtain a mass correction
\begin{equation}
1216 GeV = 7 m_t \rightarrow 
5 m_t =
875 GeV  
\end{equation}
we should seek to understand:
 How 
can this be understandable 
physically? Well, one would say,
that when the top and anti top 
particles are present in the bag,
then they interact with their 
neighbors by means of gluon 
exchange and eaten Higgs exchange,
and thus the energy density of a 
bag filled with top and anti tops
has indeed a reduced energy 
density. It would be nice to 
check that this reduction can be 
so large as we used above.

If we consider the situation 
inside the ``bag'' where the 
Higgs mass is effectively zero,
the mass of the weak gauge bosons 
must also be zero. So as long 
as we consider this ``inside'' 
the top and anti top quarks 
are attrackted by exchange of 
these gauge bosons as if they 
were massless. Also here the mass
plays no role and there is 
actually here no essential 
difference between the left b 
and the left top. So in the 
``inside'' region there should 
ideally be about equally many 
left handed top and left handed 
bottom. But right ones are only 
top-right, because the right 
bottom is in our approximation 
totally decoupled. Very crudely 
we might say that a replacement 
of half the amount of the left 
top-quarks  by bottoms means, 
that the number of quarks present
per volume unit up to a given 
energy hight gets increased by 
a factor 3/2. If we thus want to
have 12 quarks, this would in 
a calculation, in which we did 
not have this effect of the W and
Z  included, mean that we only 
should require  place for 
12:(3/2) =
8 quarks instead if using our 
calculations above. In fact we 
could claim for each 12 top or anti top, 
3 could hide as left 
bottom quarks. Crudely we 
could for a given bag size and 
given number of quarks decrease 
the needed kinetic energy per quark by a 
factor corresponding to that the 
quark-particle density could be 
decreased by the factor 
3/2. 

We might 
therefore instead of including 
the effect as we did by changing 
the effective bag-constant take 
instead an effective number 8 
for the number of quarks in the 
bound state. Now above   
we found without the corrections
\begin{equation}
M_{without \; correction}=7.03m_t
= 
1216 GeV.   
\end{equation}
In the crude thinking that the 
term with the kinetic energy 
dominated and that this term 
depends crudely proportionaly
to 3/4'th power of the number 
of particles 12. So if we reduce  this 
12 to 8 then the mass 
of the bound state should go 
down by the factor $(12/8)^{3/4}
=1.36$, and thus we would in this
way get a mass around 
\begin{equation}
7.03m_t/1.36
=
1216/1.36 GeV =5.17 m_t =897 GeV.
\end{equation}  
This bound state mass to be fully
corrected should still be 
corrected for the gluon 
contribution to the atracktion. 

Above we saw that the gluon 
correction were about a factor 
 1.523 counted in the $g_t2$. 
Saying e.g. that the binding 
energy in the Bohr atom goes as 
the coupling to the fourth power,
the binding should be increased 
by  $1.523^2 =2.32$ due to the 
gluon exchange. 

On a logarithmic scale this 
correction factor 1.523 is 
$\frac{\ln( 1.523)}
{\ln (1.523 *2)} =0.377 $, so 
that so to speak 37.7 \% of the 
correction    
$1216 GeV = 7 m_t \rightarrow 
5 m_t =
875 GeV$  is due to the 
gluons, while the remaining 
62.3 \% is due to the eaten Higgs
effect, which we have just 
replaced by its effect of 
replacing some of the top or anti
topquarks by left handed bottom quark or 
antibottom quarks. Crudely
we would estimate that the 
correction due to the gluons in 
the mass for the bound state 
which we found were a factor 
$(\frac{875}{1216})^{.377} =0.88 $
and thus we should to correct for
also the gluon effect diminish 
the 897 GeV just obtained doing 
only a replacement for the eaten Higgs correction by further 15 \%.
Thus we get the new estimate 
\begin{equation}
m_{alternative \; estimate}= .88*897 GeV = 792 GeV
\end{equation}       
\section{Conclusion}
\label{conclusion}
We have made a crude estimate of 
the mass of the bound state of 
6 top + 6 anti top quarks, about 
which we have long speculated,
that it is very strongly bound, 
to be crudely 
\begin{equation}
m_{bound \; state} = 
875 GeV 
\hbox{ or } 
792 GeV \pm \hbox{say 40 \%}.   
\end{equation}
Our method were mainly a 
bag-model estimation, in which 
the bag meant a region, where the
Higgs field were reduced to 
$\sim 0$.

The greatest importance of this 
estimate is, that it remarkably 
coincides with {\em two} earlier 
calculations {\em based on the 
assumption, that two speculated 
vacua potentially existing in 
pure Standard Model should have 
degenerate energy densities(= 
cosmologogical constants).} In 
fact the present author recently
found \cite{mass} based on a 
type of 
calculation first developped in 
the work with C.D. Froggatt
\cite{Tunguska} that the mass 
of the bound state needed for 
the degeneracy of the speculated 
phase with a condensate or at
least a higher concentration of 
these bound states with the 
present vacuum   
were
\begin{eqnarray}
m_{from \; ``condensate'' vacuum} & = & \frac{12 m_t}{3} = 4 m_t= 
692 GeV.
\end{eqnarray}
The second estimate of the bound 
state mass agreeing remarkably 
well
with the present calculation 
were in collaboration with Das 
and Laperashvili 
\cite{LNvacuumstability} and 
based on the requirement that 
there should be a vacuum - which 
we  call
``the High field vacuum'' - 
for the  Higgs field being of 
the order of $10^{18} GeV$ 
having 
with high accuracy very small 
cosmological constant or energy 
density like the present vacuum.
This is equivalent to the 
requirement that the instability 
of the present vacuum seemingly 
resulting in pure Standard Model
\cite{Deg} by means of our 
speculated bound state just gets 
corrected to be almost exactly 
on the border line of stability,
and it leads to the mass for the 
bound state  
\begin{eqnarray}
m_{bound \; state }(\hbox{from ``high field vacuum''})
&\approx & 850 GeV\pm 30 \% 
\hbox{with $\sim 2$}\\
m_{bound \; state}(\hbox{from ``high field vacuum''})
&\approx & 710 GeV\pm 30 \% 
\hbox{without $\sim 2$}.
\end{eqnarray}
The two calculations cited here 
just deviate by including 
crudely(``with $\sim$ 2'') or not 
including(``without $\sim$ 2'') 
some higher diagrams 
for the bound state causing a 
correction in the value for the 
effective self-coupling 
$\lambda$ of the 
Higgs to be used for getting the 
Higgs mass as to be observed.

Summarizing we have estimated
-although very crudely only -
the mass of the bound state in 
{\em three a priori quite different 
ways} as put in this table:

\begin{center}
\begin{tabular}{|c|c|c|c|c|}
\hline
Used &mass&In $m_t$ units& Deviation from average& Guessed Uncertainty\\
\hline
Bag-model&830 GeV&$\sim$ 5&3 \%&40 \%\\
$\Lambda_{present} =
\Lambda_{condensate}$& 690 GeV& 4
& (-)8 \%& 40 \%\\
$\Lambda_{present}= 
\Lambda_{high \; field}$& 780 GeV& $4._5$&4 \%& 30 \%\\
\hline
Average& 770 GeV& $4._3$&&21 \% \\
\hline  
\end{tabular}
\end{center}

The agreement of these mass 
estimates with each other is too
good compared to the guessed 
uncertainties of our calculations,
but the latter was not made 
carefully. If this agreement 
is taken seriously, it means that 
the degeneracies of the vacua 
as implicated by our principle 
``Multiple Point Principle'' 
could be claimed to have been 
tested by direct calculation using
the parameters of the pure 
Standard Model. 

That would then mean that we 
would have derived:

\begin{itemize}
\item The validity in the case 
of the three vacua suggested for 
the pure Standard Model the 
{\em new law of nature MPP} !
\item It would be suggested that 
no new physics should come in to 
disturb the Standard Model more 
than to not disturb the energy 
differences to our accuracy 
relative to those in the pure 
Standard Model.
\item The bound state - 
that shall do the job - should
really exist!
\item The mass of it must be our estimated 
770 GeV $\pm $ 19 \%.
(It is really sad for our picture
that the enhancement known as 
$F(750)$ with just the right mass
and decaying into two gammas, 
were washed out so that no 
statistics remains supporting 
it! We miss it!)  
\end{itemize}

It should be stressed that in 
principle - i.e. if we can 
perform non-perturbative 
calculations sufficiently 
accurately - we should be able
to simply calculate, if there 
exist the above posulated vacua.
Well, the ``high field one'' 
confrontation with MPP requires 
that  one
presupposes that the Standard 
Model to be valid to sufficient 
accuracy all the way up to about
$10^{18} GeV$. Even for the 
``condensate vacuum'' one could 
imagine that new physics might 
modify our calculations, but LHC
has already put severe limits 
telling, that there is no 
new physics up to a scale of 
the order of one TeV and thus 
our proposed bound state of a 
mass of the order of 
770 GeV is
expected to be not very sensitive
to at present acceptable new 
physics. Thus improved 
calculational methods for 
non-perturbative effects, 
especially strongly bound states,
should possibly 
 rather independently of 
possible new physics 
   be able to settle, if our bound
state of a mass in the range near
the value of 750 GeV really 
exists or not.

\section*{Acknowledgement}
I want to thank the Niels Bohr Institute 
for allowing me to stay as emeritus. 
Also I thank my near collaborators on 
the subject of the present article 
Larisa Laperashvili, Colin D. Froggatt
and Chitta Das for the collaboration, 
without which the present article 
could never have come to being.
Further I thank for the comments I have 
got at Mittag Lefler Insitute, the 
Bled workshop ``Beyond the Standard 
Models'' and The Corfu Conference 2016, 
where I have had some 
presentation of essentially the present 
work.   

Astri Kleppe even presented our 
``Mutiple Point Principle'' for the 
workshop in Bled.

\end{document}